\documentclass{kapproc} 

\RequirePackage{graphicx}%
\RequirePackage{epsf}%
\input epsf

\upperandlowercase
\let\footnote\savefootnote
\let\footnotetext\savefootnotetext 
 
\kluwerbib 

\def\apj{ApJ}

\def\araa{ARA\&A}
\def\mnras{MNRAS}

\def\pasp{PASP}

\def\s2n{S^{\prime}/N}

\begin{document}
\articletitle{The Stellar IMF as a Property of Turbulence}

\chaptitlerunninghead{The Stellar IMF as a Property of Turbulence}

\author{Paolo Padoan\altaffilmark{1}, \AA ke Nordlund\altaffilmark{2}}

\affil{
\altaffilmark{1}Department of Physics, University of California, San Diego, CASS/UCSD 0424, 9500 Gilman Drive, La Jolla, CA 92093-0424\\
\altaffilmark{2}Astronomical Observatory / NBIfAFG, Juliane Maries Vej 30, DK-2100, Copenhagen, Denmark
}

\email{ppadoan@ucsd.edu,aake@astro.ku.dk}

\begin{abstract}
We propose to interpret the stellar IMF as a property of the 
turbulence in the star--forming gas. Gravitationally unstable
density enhancements in the turbulent flow collapse and form stars. 
Their mass distribution can be derived analytically from the power 
spectrum of the turbulent flow and the isothermal shock jump conditions 
in the magnetized gas. For a power spectrum index $\beta=1.74$, 
consistent with Larson's velocity dispersion--size relation as well 
as with new numerical and analytic results on supersonic turbulence, 
we obtain a power law mass distribution of dense cores with a slope 
equal to $3/(4-\beta) = 1.33$, consistent with the slope of Salpeter's 
stellar IMF. Below one solar mass, the mass distribution flattens and turns 
around at a fraction of a solar mass, as observed for the stellar IMF 
in a number of stellar clusters, because only the densest cores are 
gravitationally unstable. The mass distribution at low masses is 
determined by the Log--Normal distribution of the gas density. The 
intermittent nature of this distribution is responsible for the generation 
of a significant number of collapsing cores of brown dwarf mass.
\end{abstract}

\section{Introduction}
The origin of the stellar initial mass function (IMF) is a fundamental 
problem in astrophysics because the stellar IMF determines photometric 
properties of galaxies and the dynamical and chemical evolutions of their
interstellar medium. In this contribution we address the relation between 
statistical properties of turbulence and the origin of the stellar IMF
as discussed in Padoan \& Nordlund (2002, 2004).
The main result of these works is that the power law slope, $s$, of the stellar
IMF measured by \cite{Salpeter55} is the consequence of the turbulent 
nature of the star--forming gas and is directly related to the turbulent
power spectrum slope, $\beta$, $s=3/(4-\beta)$. From this point of view it 
is not surprising that the origin of Salpeter's result has remained 
mysterious for half a century, as our understanding of turbulence 
has not improved much since the seminal work by \cite{Kolmogorov41}. 
The situation has changed during the last decade, because
ever increasing computer resources have recently allowed significant
progress in both fields of turbulence and star formation.  
 
Since Salpeter's work, the stellar IMF has been measured successfully
in many systems. We now know that the IMF in young clusters 
(e.g. Chabrier 2003) reaches a maximum at a fraction of a solar 
mass, and then turns around with a relative abundance of brown dwarfs
(BDs) that may vary from cluster to cluster (e.g. Luhman et al. 2000).
The work we present here explains also this feature of the IMF as a 
consequence of the supersonic turbulence in the star--forming gas. 

Using the properties of supersonic turbulence we have derived from 
recent numerical simulations, we predict the stellar IMF essentially 
without free parameters. This predicted IMF is shown to depend on the 
rms Mach number, mean density and temperature of the turbulent flow. 
It is also shown to agree well with Salpeter's IMF for large stellar 
masses and with the low mass IMF derived for young stellar clusters. 
Our view of the IMF as a natural property of supersonic turbulence 
in the magnetized and isothermal star--forming gas provides an 
explanation for the origin of BDs as well. According to 
this picture, BDs may be formed in the same way as 
hydrogen--burning stars.

\section{Statistics of Supersonic MHD Turbulence}

The velocity power spectrum in the inertial range of turbulence (between
the scales of energy injection and viscous dissipation) is a power law,
$E_v(k)\propto k^{-\beta}$, where $k$ is the wave--number. The spectral 
index is $\beta\approx5/3$ for incompressible turbulence (Kolmogorov 1941), 
and $\beta\approx2$ for pressureless turbulence (Burgers 1974; Gotoh \& Kraichnan1993).
In recent numerical simulations of isothermal, super--Alfv\'{e}nic and highly 
supersonic magneto--hydrodynamic (MHD) turbulence we have obtained a power 
spectrum intermediate between the Burgers and the Kolmogorov power spectra 
(Boldyrev, Nordlund \& Padoan 2002) and consistent with the prediction by 
\cite{Boldyrev02}, $E(k)\propto k^{-1.74}$.

\begin{figure}[ht]
\centerline{
\epsfxsize=10cm \epsfbox{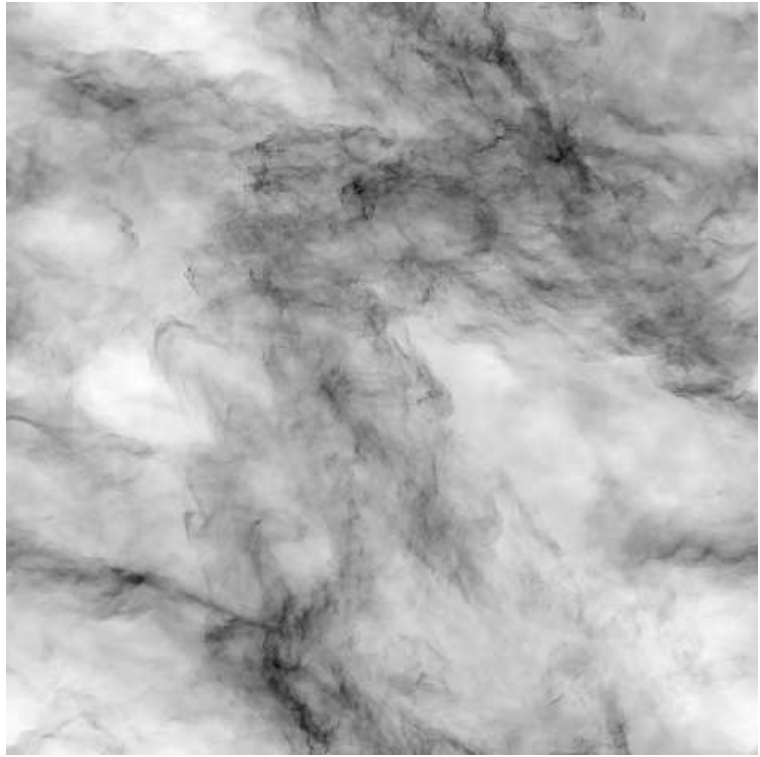}
}
\caption[]{Projected density field from a numerical simulation of isothermal,
supersonic hydrodynamic turbulence with an effective resolution of $1024^3$ 
computational zones (Kritsuk, Padoan \& Norman, in preparation). The contrast 
has been reduced in order to show details of the low density regions.}
\label{fig0}
\end{figure}

The probability density function (PDF) of gas density in isothermal 
turbulent flows is well approximated by a Log--Normal distribution  
with moments depending on the rms Mach number of the flow 
(Nordlund \& Padoan 1999; Ostriker, Gammie \& Stone 1999). The density
structure is characterized by a complex system of interacting shocks 
resulting in a fractal network of dense cores, filaments, sheets and 
low density ''voids'', with a large density contrast. Most of the mass 
concentrates in a small volume fraction (according to the Log--Normal PDF), 
a manifestation of the intermittent nature of the turbulence.
An example of a projected turbulent density field is shown in Figure~\ref{fig0}.
\section{From Kolmogorov to Salpeter: The Mass Distribution of Collapsing Cores}

A simple model of the expected mass distribution of dense cores
generated by supersonic turbulence has been proposed in
\cite{Padoan+Nordlund02imf}. The model is based on two assumptions:
i) The power spectrum of the turbulence is a power law; ii) the
typical size of a dense core scales as the thickness of the postshock
gas. The first assumption is a basic result for turbulent flows and
holds also in the supersonic regime (Boldyrev et al. 2002).
The second assumption is suggested by the fact that postshock condensations
are assembled by the turbulent flow in a dynamical time.
Condensations of virtually any size can therefore be formed, 
independent of their Jeans' mass.

With these assumptions, together with the jump conditions for MHD
shocks, the mass distribution of dense cores can be related to the 
power spectrum of turbulent velocity, $E_v(k)\propto k^{-\beta}$:
\begin{equation}
N(m)\,{\rm d}\ln m\propto m^{-3/(4-\beta)}{\rm d}\ln m ~.
\label{imf}
\end{equation}
If the turbulence spectral index $\beta$ is taken from the
analytical prediction by \cite{Boldyrev02}, which
is consistent with the observed velocity dispersion--size Larson 
relation (Larson 1979, 1981) and with our numerical results 
(Boldyrev et al. 2002), then $\beta \approx 1.74$ and the 
mass distribution is $N(m)\,{\rm d}\ln m\propto m^{-1.33}{\rm d}\ln m$,
almost identical to Salpeter's stellar IMF (Salpeter 1955).
The exponent of the mass distribution is rather well constrained,
because the value of $\beta$ for supersonic turbulence cannot be
smaller than the incompressible value, $\beta= 1.67$ (slightly larger
with intermittency corrections), and the Burgers case, $\beta=2.0$.
As a result, the exponent of the mass distribution is predicted to be 
within the range of values of 1.3 and 1.5. 

While massive cores are usually larger than their critical Bonnor--Ebert mass,
$m_{\rm BE}$, the probability that small cores are dense enough to collapse is
determined by the statistical distribution of core density. 
In order to compute this collapse probability for small cores, we assume 
i) the distribution of core density can be approximated by the 
\begin{figure}[ht]
\centerline{
\epsfxsize=6cm \epsfbox{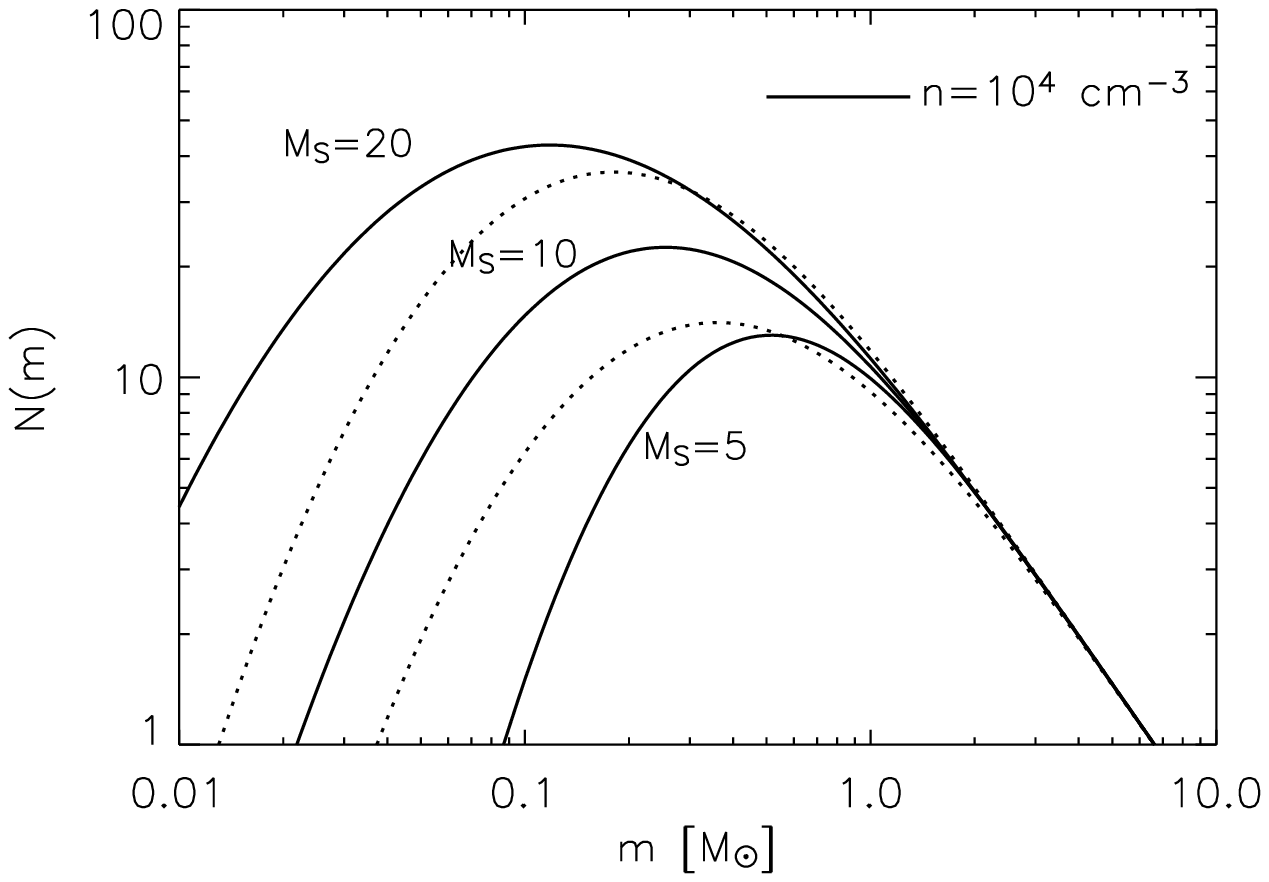}
\epsfxsize=6cm \epsfbox{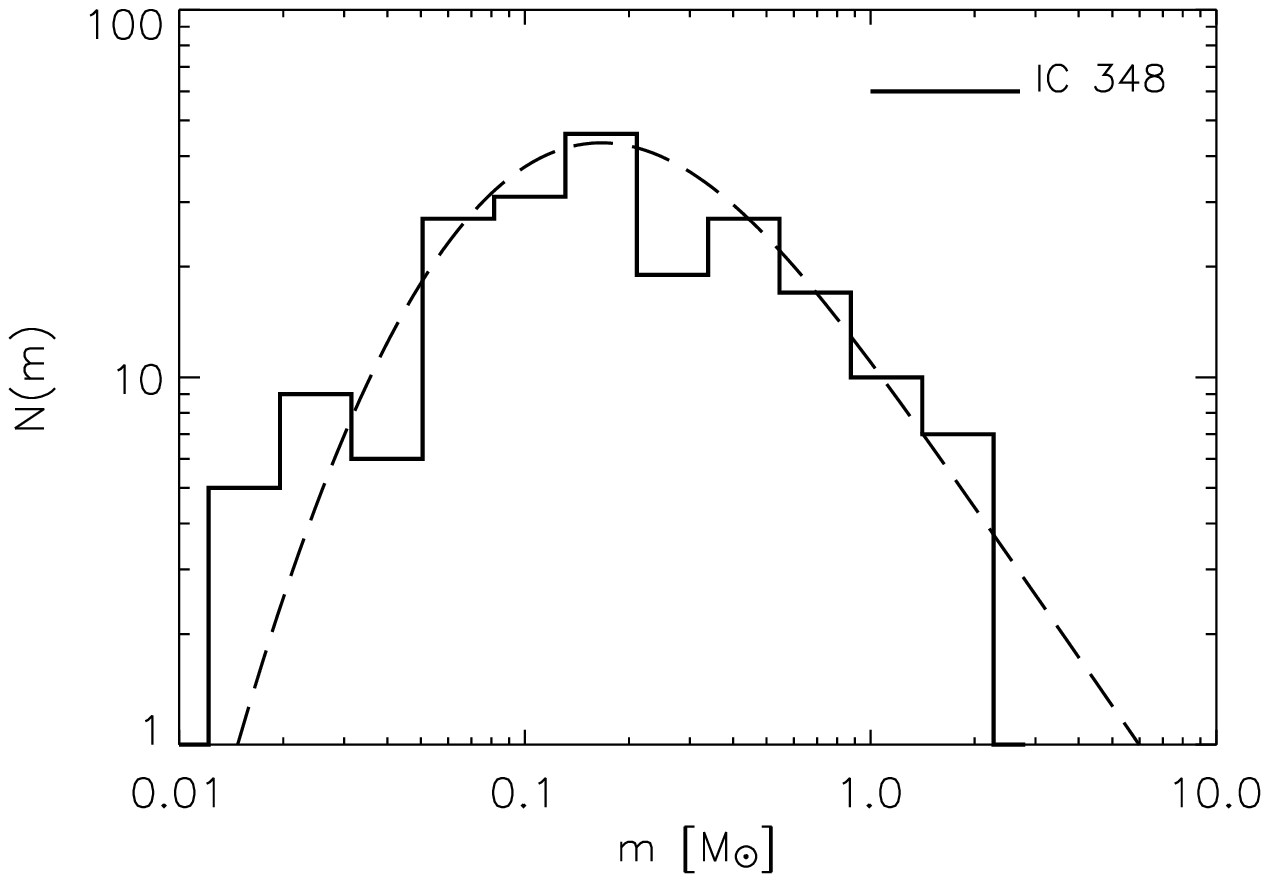}
}
\caption[]{Left panel: Analytical mass distributions computed
for $\langle n\rangle=10^4$~cm$^{-3}$, $T=10$~K and 
for three values of the sonic rms Mach number, $M_{\rm S}=5$, 10 and 20
(solid lines). The dotted lines show the mass distribution for 
$T=10$~K, $M_{\rm S}=10$ and $\langle n\rangle=5\times10^3$~cm$^{-3}$
(lower plot) and $\langle n\rangle=2\times10^4$~cm$^{-3}$ (upper plot).
Right panel: IMF of the cluster IC 348 in Perseus obtained by 
\cite{Luhman+2003} (solid line histogram) and theoretical IMF computed for 
$\langle n\rangle=5\times 10^4$~cm$^{-3}$, $T=10$~K and 
$M_{\rm S}=7$ (dashed line).}
\label{fig1}
\end{figure}
Log--Normal PDF of gas density and ii) the core density and mass are 
statistically independent. Because of the intermittent
nature of the Log-Normal PDF, even very small (sub--stellar) cores
have a finite chance to be dense enough to collapse.
Based on the first assumption, we can compute the distribution of the 
critical mass,  $p(m_{\rm BE})\,d m_{\rm BE}$, from the Log--Normal
PDF of gas density assuming constant temperature (Padoan, Nordlund \& Jones 1997).
The fraction of cores of mass $m$ larger than their critical
mass is given by the integral of $p(m_{\rm BE})$ from 0 to
$m$. Using the second assumption of statistical independence of core density 
and mass, the mass distribution of collapsing cores is
\begin{equation}
N(m)\, {\rm d}\ln m\propto m^{-3/(4-\beta)}\left[\int_0^m{p(m_{\rm BE}){\rm d}m_{\rm BE}}\right]\,{\rm d}\ln m ~.
\label{imfpdf}
\end{equation}
This mass distribution is a function of the rms Mach number, mean density and 
temperature of the turbulent flow, as these parameters enter the PDF of gas 
density and thus $p(m_{\rm BE})$. Figure~\ref{fig1} (left panel) shows five 
mass distributions computed from equation (\ref{imfpdf}) with three different 
values of the sonic rms Mach number and two different values of density. 
\cite{Padoan+Nordlund04bd} have suggested that BDs may originate from 
the process of turbulent fragmentation like hydrogen--burning stars.
This is illustrated by the analytical IMF in the left panel of Figure~\ref{fig1}, 
which shows a relatively large abundance of BDs is predicted for sufficiently
large values of the mean density or rms Mach number.

The IMF of the cluster IC 348 in Perseus, obtained by \cite{Luhman+2003}, is 
plotted in the right panel of Figure~\ref{fig1} (solid line histogram). 
The IMF of this cluster has been chosen for the comparison with the theoretical 
model because it is probably the most reliable observational IMF including both 
BDs and hydrogen burning stars. In Figure~\ref{fig1} we have also plotted the 
theoretical mass distribution computed for $\langle n\rangle=5\times 10^4$~cm$^{-3}$, 
$T=10$~K and $M_{\rm S}=7$. These parameters are appropriate
for the central $5\times 5$~arcmin of the cluster ($0.35\times 0.35$~pc),
where the stellar density corresponds to approximately $2\times 10^4$~cm$^{-3}$.
The figure shows that the theoretical distribution of collapsing 
cores, computed with parameters inferred from the observational data, 
is roughly consistent with the observed stellar IMF in the cluster IC 348. 
Similar IMFs have been obtained for several other young clusters (Chabrier 2003).

\section{Conclusions}    
 
We have proposed to explain the stellar IMF as a property of supersonic 
turbulence. This scenario is very different from previous theories of star 
formation (see Shu, Adams \& Lizano 1987), where it is assumed that stars of small 
and intermediate mass are formed from sub--critical cores evolving 
quasi--statically, on the time--scale of ambipolar drift. These theories
do not account for the ubiquity of turbulence in star--forming clouds
and therefore ignore the effect of turbulence in the fragmentation process.

A naive interpretation of our results may lead to the conclusion that 
the stellar IMF at large masses should be a universal power law, with
a slope very close to Salpeter's value. The statistics of turbulence
controlling the origin of the stellar IMF are indeed universal (they
depend on {\it flow} properties such as the rms Mach number, not
{\it fluid} properties, and are insensitive to initial conditions
that are soon ``forgotten'' due to the chaotic nature of the turbulence). 
However, such statistics are derived from ensemble averages.
According to our derivation, massive stars originate from shocks
on relatively large scales. In any given system (for example a molecular cloud)
the number of large scale compressions (thus the number of massive stars)
is relatively small. Because of the small size of the statistical sample
and because of the intermittent nature of the turbulence, large deviations 
from Salpeter's IMF are possible in individual systems. Salpeter's IMF is 
therefore a universal property of supersonic turbulence, but it is
not necessarily reproduced precisely in every star--forming region.


\begin{chapthebibliography}{}

\expandafter\ifx\csname natexlab\endcsname\relax\def\natexlab#1{#1}\fi

\bibitem[{{Boldyrev} (2002)}]{Boldyrev02}
{Boldyrev}, S. 2002, \apj, 569, 841

\bibitem[{{Boldyrev} {et~al.} (2002)}]{Boldyrev+02scaling}
{Boldyrev}, S., {Nordlund}, {\AA}., \& {Padoan}, P. 2002, \apj, 573, 678

\bibitem[{Burgers (1974)}]{Burgers74}
Burgers, J.~M. 1974, in The Nonlinear Diffusion Equation (Reidel, Dordrecht)

\bibitem[{{Chabrier} (2003)}]{Chabrier03}
{Chabrier}, G. 2003, \pasp, 115, 763

\bibitem[{Gotoh \& Kraichnan (1993)}]{Gotoh+Kraichnan93}
Gotoh, T. \& Kraichnan, R.~H. 1993, Phys. Fluids, A, 5, 445

\bibitem[{Kolmogorov (1941)}]{Kolmogorov41}
Kolmogorov, A.~N. 1941, Dokl. Akad. Nauk. SSSR, 30, 301

\bibitem[{{Larson} (1979)}]{Larson79}
{Larson}, R.~B. 1979, \mnras, 186, 479

\bibitem[{{Larson} (1981)}]{Larson81}
---. 1981, \mnras, 194, 809

\bibitem[{{Luhman} {et~al.} (2000)}]{Luhman+2000}
{Luhman}, K.~L., {Rieke}, G.~H., {Young}, E.~T., {Cotera}, A.~S., {Chen}, H.,
  {Rieke}, M.~J., {Schneider}, G., \& {Thompson}, R.~I. 2000, \apj, 540, 1016

\bibitem[{{Luhman} {et~al.} (2003)}]{Luhman+2003}
{Luhman}, K.~L., {Stauffer}, J.~R., {Muench}, A.~A., {Rieke}, G.~H., {Lada},
  E.~A., {Bouvier}, J., \& {Lada}, C.~J. 2003, \apj, 593, 1093

\bibitem[{{Nordlund} \& {Padoan} (1999)}]{Nordlund+Padoan99}
{Nordlund}, {\AA}.~K. \& {Padoan}, P. 1999, in Interstellar Turbulence, 218

\bibitem[{{Ostriker} {et~al.} (1999)}]{Ostriker+99}
{Ostriker}, E.~C., {Gammie}, C.~F., \& {Stone}, J.~M. 1999, ApJ, 513, 259

\bibitem[{{Padoan} \& {Nordlund} (2002)}]{Padoan+Nordlund02imf}
{Padoan}, P. \& {Nordlund}, {\AA}. 2002, \apj, 576, 870

\bibitem[{{Padoan} \& {Nordlund} (2004)}]{Padoan+Nordlund04bd}
---. 2004, ApJ, in press (see also astro-ph/0205019)

\bibitem[{Padoan {et~al.} (1997)}]{Padoan+97imf}
Padoan, P., Nordlund, {\AA}., \& Jones, B. 1997, MNRAS, 288, 145

\bibitem[{{Salpeter} (1955)}]{Salpeter55}
{Salpeter}, E.~E. 1955, \apj, 121, 161

\bibitem[{Shu {et~al.} (1987)}]{Shu+87}
{Shu}, F.~H., {Adams}, F.~C. \& {Lizano}, S., 1987, \araa, 25, 23 


\end{chapthebibliography}
\end{document}